\documentclass[referee]{aa} % for a referee version
\usepackage{natbib}
\usepackage{graphicx}
\usepackage{txfonts}
\usepackage{epsfig}
\usepackage{times}
\usepackage{rotating}
\usepackage{setspace}
\def\gtsim{\mathrel{\spose{\lower.5ex \hbox{$\mathchar"218$}}
     \raise.4ex\hbox{$\mathchar"13E$}}}
\def\ltsim{\mathrel{\spose{\lower.5ex\hbox{$\mathchar"218$}}
     \raise.4ex\hbox{$\mathchar"13C$}}}
\def\spose#1{\hbox to 0pt{#1\hss}}

\def\apjl{ApJ}                % Astrophysical Journal, Letters
\def\apjs{ApJS}               % Astrophysical Journal, Supplement
\def\ao{Appl.~Opt.}           % Applied Optics
\def\aap{A\&A}                % Astronomy and Astrophysics
\def\ot{[\ion{O}{iii}]$\lambda\lambda4959, 5007$}
\def\oiiiq{[\ion{O}{iii}]$\lambda4959$}
\def\oiiic{[\ion{O}{iii}]$\lambda5007$}
\def\od{[\ion{O}{ii}]$\lambda3727$}
\def\Hb{${\rm H}{\small{\beta}}$}
\def\Ha{${\rm H}{\small{\alpha}}$}
\def\nii{[\ion{N}{ii}]$\lambda6583$}
\def\rdt{{\it R}$_{23}$}
\def\otd{{\it O}$_{32}$}
\def\oh{$12+\rm log(O/H)$}
\begin{document}

\title{Gas-phase metallicity of 27 galaxies at intermediate redshift\thanks{Based 
    on observations made with ESO Telescopes at the La Silla-Paranal Observatory 
    under programmes 075.B-0794 and 077.B-0767.}}

\author{L. Morelli\inst{1,2}
        \and
        V. Calvi\inst{1,3}
        \and
        A. Cardullo\inst{1}
        \and
        A. Pizzella\inst{1,2}
        \and
	E. M. Corsini\inst{1,2}
        \and
        E. Dalla Bont\`a\inst{1,2}
	}

\institute{Dipartimento di  Fisica e Astronomia ``G. Galilei", Universit\`a di Padova,
              vicolo dell'Osservatorio 3, I-35122 Padova, Italy\\
              \email{lorenzo.morelli@unipd.it}
              \and
              INAF-Osservatorio Astronomico di Padova,
              vicolo dell'Osservatorio~2, I-35122 Padova, Italy\\
              \and
              Space Telescope Science Institute, 3700 San Martin Drive,
              Baltimore, MD 21218, USA\\
                          }

\date{\today}

\abstract{} {The purpose of this work is to make available new
  gas-phase oxygen abundance measurements for a serendipitous sample
  of 27 galaxies with redshifts $0.35 \leq z \leq 0.52$. }  {We measured
  the equivalent widths of the \od , \Hb , and \ot\/ emission lines
  observed in the galaxy spectra obtained with the Visible
  Multi-Object Spectrograph mounted at the Very Large Telescope.  For
  each galaxy, we derived the metallicity-sensitive emission-lines
  ratio \rdt, ionization-sensitive emission-lines ratio \otd, and
  gas-phase oxygen abundance \oh\/. } { The values of the gas-phase oxygen
  abundance \oh\/ that we obtained for all the sample galaxies are consistent
  with previous findings for galaxies at intermediate redshift. }{}

\maketitle

\section{Introduction}
\label{sec:introduction}

The gas-phase metallicity is one of the most important observational
diagnostics of the current evolutionary state of galaxies.  Accurate
gas-phase metallicity measurements are crucial for investigating the
interplay between fundamental processes, such as star formation, gas
accretion, gas flows, and supernova-driven winds occurring during the
life of galaxies \citep[e.g.,][]{Moustakas2011, cresci2012,
  sommariva2011}.

The galaxy gas-phase metallicity correlates with many properties, such
as the star formation rate \citep{mannucci2010, laralopez2010},
morphological type \citep{edmundspagel1984}, surface mass density
\citep{ryder1995, garnett1997}, and maximum rotation velocity
\citep{dalcanton2007}. In particular, the metallicity $Z$ most strongly
correlates with the mass $M$ and $B$-band luminosity $L$ of the galaxy
\citep{mcclure1968, skillman89, richer95}.
Adopting the Sloan Digital Sky Survey (SDSS, \citealt{DR7}),
\cite{tremonti2004} and \cite{gallazzi2005} derived the $M-Z$ and
$L-Z$ relations in the local Universe for a large sample of galaxies
on the basis of the gas-phase metallicity. These relations are very
tight with a scatter of less than 0.1 dex and show that the more
luminous galaxies have higher gas metallicities than their fainter
counterparts, which has a great impact on the theoretical models of galaxy
formation.

Another crucial piece of information to help us to understand the assembly
history of galaxies comes from the study of the evolution of the $M-Z$
and $L-Z$ relations across cosmic time. 

This topic has been investigated by measuring the gas-phase metallicity
of galaxies as a function of redshift up to $z\sim3.5$.
While these relations show clear evidence of galactic evolution at high $z$, because at
a given mass higher-$z$ galaxies have lower metallicities
\citep{shapley2004, erb2006, maiolino2008, mannucci2009, rodretal2012}, at
intermediate redshift ($z<1$) it is still debated whether there has been
any evolution.  \cite{carollo2001} and \citet{kobulnicky2004} found
that the $L-Z$ relation at intermediate redshift is consistent with
the local one. In contrast, \cite{maier2005}, \cite{savaglio2005},
and \cite{zahid2011} measured lower metallicities for
intermediate-redshift objects than for local galaxies, supporting a
scenario in which the $L-Z$ and $M-Z$ relations also evolve over the
range of redshifts between $0.5$ and $1$.

In this research note, we therefore present new measurements of the
equivalent widths for the \od\/, \Hb\/, \oiiiq\/, and \oiiic\/
emission-lines and present our derived metallicity-sensitive emission
lines ratios \rdt, ionization-sensitive emission-lines ratios \otd,
and gas-phase oxygen abundances \oh\/ for 27 intermediate-redshift
($0.35 \leq z \leq 0.52$) galaxies. The data that we analised for all
the sample galaxies are available from the SDSS archive, hence our
measurements are a valuable supplementary resource for the
astronomical community.

\section{Observations and data reduction}
\label{sec:data}

As often happens, the data acquired for a particular aim can
potentially contain information useful for different purposes. For
this work, we used observations carried out in 2006 at the Very
Large Telescope (VLT) of the European Southern Observatory (ESO) at
Paranal Observatory using the VIsible Multi-Object Spectrograph
(VIMOS).  The main goal of these observations was to find satellites
in the region of $\sim500 \times 500~ {\rm kpc}^2 $ surrounding seven
nearby isolated spiral galaxies \citep{yegorova2011}.

VIMOS was used in Multi Object Spectroscopy (MOS) mode with the grism
``HR orange'' and the order sorting-filter GG435. The grism is
characterized by a reciprocal dispersion of ${ 0.6~ \rm \AA}\; {\rm
  pixel}^{-1}$ and a spectral resolution of $R=2150$ for the adopted
$1\farcs0$ slit. In the high-resolution configuration that we used for
the observations, the spectral range strongly depends on the position
of the slit in the field of view varying from $4550-6950~ \AA$ to
$6000-8400~ \AA$.  The full dataset includes MOS observations of 7
different fields for a total exposure time of $\sim 20$ hours. This
granted a typical exposure time of $\sim 3$ hours for each target
field. Each of the four quadrants of VIMOS observed $\sim 50$ objects
so as to have $\sim 200$ galaxies within each field of view of
$16\farcm0 \times 18\farcm0$.

 The observed sample consists of 1450 objects with SDSS $r-$band
  magnitudes ranging from 17 mag to 24 mag and a modal value of 22 mag. As
  expected, only a small fraction of the target galaxies were found to be
  satellites. We focused on the remaining 1347 background and
  foreground objects to measure their gas-phase metallicities.
Their spectra were bias-subtracted, corrected for flat-field effect, cleaned of
cosmic rays, and wavelength-calibrated by means of the ESO VIMOS data
reduction pipeline.  A more detailed description of the data reduction
is reported in \cite{yegorova2011}.

\section{Sample selection}
\label{sec:data}

To estimate the gas-phase metallicity, it is necessary to know the
electron temperature provided by the comparison between auroral and
nebular emission lines \citep{osterbrock1989}.  However, since auroral
lines are generally weak, it is difficult to observe them in distant
galaxies. To overcome this issue, empirical relations between the
intensity of the nebular lines and metallicity have been used to
derive the gas-phase metallicity. The most widely used metallicity
indicator is the \rdt\/ emission lines ratio parameter
\citep{pagel1979} defined as
\begin{equation}
{{\it R}_{23}}=\frac{[\ion{O}{ii}]\lambda3727+[\ion{O}{iii}]\lambda4959+
[\ion{O}{iii}]\lambda5007}{\rm H{\small{\beta}}}.
\label{eq:r23}
\end{equation}

For our analysed dataset the signal-to-noise ratio ($S/N$) of the spectra allowed us to measure
redshifts for only 571 objects out of the observed 1347 galaxies.
We fitted all the available emission lines with Gaussians. The \od\/
doublet was modelled with two Gaussians of the same full width at
half maximum. The galaxy redshift was computed as the average of the
redshifts derived from the central wavelengths of the measured
emission lines.

\begin{figure}
\includegraphics[angle=0.0,width=0.5\textwidth]{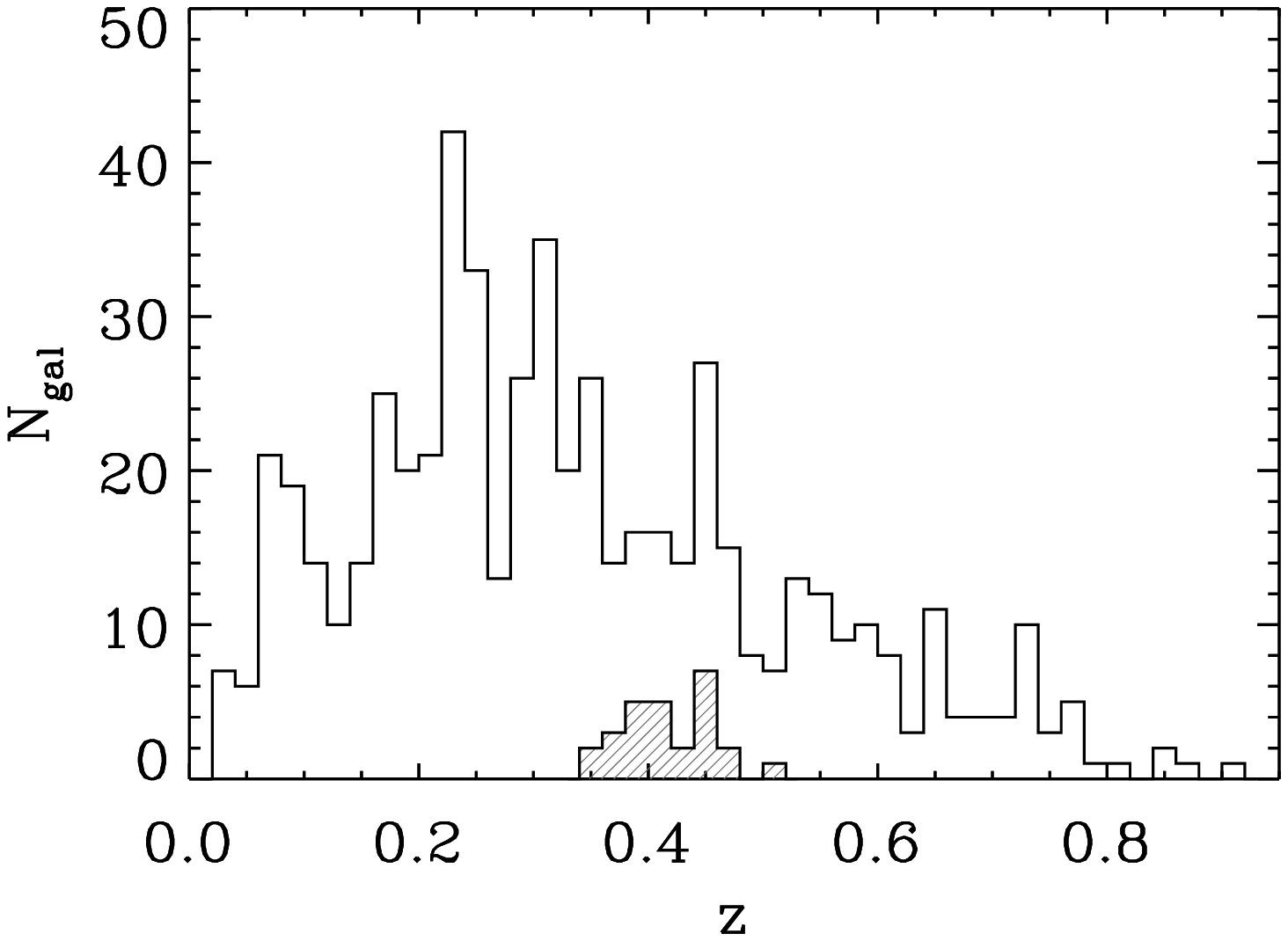}
\caption{Distribution of redshifts for the 571 galaxies with
  measured emission lines. The shaded histogram shows the distribution
  of the serendipitous sample of 27 objects presented in this work.}
\label{z_distribution}
\end{figure}

We detected 177 galaxies in the redshift range between $0.35$ and
$0.6$. However, most of their spectra did not map all the \od\/, \Hb
, and \ot\/ emission-lines we needed to compute the metallicity-sensitive
emission-lines ratio \rdt, ionization-sensitive emission lines
ratio \otd, and gas-phase oxygen abundance \oh\/, because of their
actual position in the VIMOS field of view.  All the relevant emission
lines fall in the observed spectral range only for 27 objects, which
represent our final sample of galaxies. They are listed in Table
\ref{tab:values} and the distribution of their measured redshifts is
shown in Figure \ref{z_distribution}.

%%%%%%%%%%%%%%%%%%%%%%%%%%%%%%%%%%%%%%%%%%%%%%%%%%%%%%%%%%%%%
%% FIGURA ISTOGRAMMI
%%%%%%%%%%%%%%%%%%%%%%%%%%%%%%%%%%%%%%%%%%%%%%%%%%%%%%%%%%%%%

%%%%%%%%%%%%%%%%%%%%%%%%%%%%%%%%%%%%%%%%%%%%%%%%%%%%%%%%%%%%%
%% TABELLA VALORI MISURATI - ATTENZIONE ALLA CAPTION
\begin{table*}
\centering
\caption{Measured properties of the sample galaxies.}
\begin{small}
\begin{tabular}{crrrrr}
\hline
\noalign{\smallskip}
\multicolumn{1}{c}{Galaxy name} &
\multicolumn{1}{c}{\od} &
\multicolumn{1}{c}{\Hb} &
\multicolumn{1}{c}{\oiiiq} &
\multicolumn{1}{c}{\oiiic} &
\multicolumn{1}{c}{$z$} \\
\multicolumn{1}{c}{SDSS} &
\multicolumn{1}{c}{[$\AA$]} &
\multicolumn{1}{c}{[$\AA$]} &
\multicolumn{1}{c}{[$\AA$]} &
\multicolumn{1}{c}{[$\AA$]} &
\multicolumn{1}{c}{} \\
\multicolumn{1}{c}{(1)} &
\multicolumn{1}{c}{(2)} &
\multicolumn{1}{c}{(3)} &
\multicolumn{1}{c}{(4)} &
\multicolumn{1}{c}{(5)} &
\multicolumn{1}{c}{(6)} \\
\noalign{\smallskip}
\hline
\noalign{\smallskip}
J003845.31+000006.8 &$  2.95 \pm 0.45$ &$     1.43  \pm   0.49$ &$    10.01  \pm  0.66$ &$    26.35 \pm   0.45 $&$0.451  $\\
J003838.68+000225.1 &$ 92.41 \pm 0.68$ &$    22.22  \pm   0.31$ &$    15.58  \pm  0.72$ &$    46.25 \pm   0.69 $&$0.408  $\\
J003806.59+000421.9 &$ 208.77\pm 7.16$ &$    15.75  \pm   1.07$ &$     9.98  \pm  2.50$ &$    39.45 \pm   1.30 $&$0.474  $\\
J134235.94+014424.1 &$  3.89 \pm 0.41$ &$     3.78  \pm   0.54$ &$     1.61  \pm  0.46$ &$     9.17 \pm   0.68 $&$0.411  $\\
J134229.37+015244.5 &$ 58.21 \pm 0.58$ &$    20.63  \pm   0.48$ &$    17.33  \pm  0.77$ &$    43.93 \pm   0.69 $&$0.372  $\\
J134156.84+014241.0 &$  6.09 \pm 0.47$ &$     4.79  \pm   1.57$ &$     3.74  \pm  0.87$ &$    13.11 \pm   0.82 $&$0.390  $\\
J145221.51+043448.8 &$ 23.01 \pm 0.55$ &$     9.14  \pm   0.15$ &$    -0.12  \pm  0.45$ &$     5.17 \pm   0.35 $&$0.450  $\\
J145150.75+044111.6 &$  7.09 \pm 0.37$ &$     3.23  \pm   0.83$ &$     2.74  \pm  0.78$ &$     2.76 \pm   0.40 $&$0.425  $\\
J145147.24+043654.2 &$ 30.29 \pm 1.13$ &$     6.28  \pm   1.68$ &$     4.61  \pm  1.35$ &$     6.12 \pm   1.17 $&$0.517  $\\
J145156.20+043321.7 &$  7.92 \pm 0.61$ &$     2.16  \pm   0.12$ &$     4.13  \pm  0.25$ &$    14.93 \pm   0.32 $&$0.416  $\\
J145147.77+043244.5 &$ 11.84 \pm 0.61$ &$     1.03  \pm   0.79$ &$     2.70  \pm  0.75$ &$     4.18 \pm   0.81 $&$0.414  $\\
J145149.20+043501.6 &$ 14.03 \pm 1.09$ &$     2.90  \pm   0.50$ &$     1.57  \pm  0.84$ &$     8.16 \pm   0.67 $&$0.443  $\\
J152638.18+034353.8 &$ 15.42 \pm 0.36$ &$     9.75  \pm   0.46$ &$     4.80  \pm  0.79$ &$    15.49 \pm   0.51 $&$0.407  $\\
J152630.18+035110.8 &$ 19.58 \pm 0.57$ &$     5.93  \pm   0.45$ &$     3.27  \pm  1.01$ &$     4.39 \pm   0.60 $&$0.458  $\\
J152559.83+035249.3 &$ 18.67 \pm 0.45$ &$     3.24  \pm   0.59$ &$     3.30  \pm  0.67$ &$     8.50 \pm   0.56 $&$0.475  $\\
J152606.03+034530.7 &$ 34.36 \pm 0.54$ &$     9.08  \pm   0.65$ &$     6.12  \pm  0.44$ &$    30.29 \pm   0.36 $&$0.386  $\\
J154933.50-004255.0 &$ 39.71 \pm 0.57$ &$    14.98  \pm   0.27$ &$    14.95  \pm  0.68$ &$    51.32 \pm   0.46 $&$0.358  $\\
J154931.48-003406.1 &$ 45.92 \pm 0.59$ &$    21.36  \pm   0.62$ &$    28.59  \pm  0.42$ &$   107.08 \pm   0.36 $&$0.386  $\\
J154925.24-003717.8 &$ 18.21 \pm 0.45$ &$     4.76  \pm   0.63$ &$     2.99  \pm  0.32$ &$    13.41 \pm   0.32 $&$0.386  $\\
J154914.49-004010.2 &$ 30.77 \pm 0.65$ &$     9.05  \pm   0.43$ &$     4.52  \pm  0.89$ &$    11.21 \pm   0.51 $&$0.441  $\\
J154925.12-003645.4 &$ 18.54 \pm 0.34$ &$     5.50  \pm   0.45$ &$     1.90  \pm  0.19$ &$     7.83 \pm   0.37 $&$0.445  $\\
J222013.98-074355.8 &$  6.41 \pm 0.47$ &$     2.06  \pm   0.20$ &$     1.00  \pm  0.32$ &$     1.29 \pm   0.24 $&$0.399  $\\
J222014.86-073333.7 &$ 23.12 \pm 1.01$ &$     7.10  \pm   0.88$ &$     1.51  \pm  1.52$ &$    10.07 \pm   1.57 $&$0.440  $\\
J221934.90-073825.9 &$  7.73 \pm 0.89$ &$     3.44  \pm   0.29$ &$     8.27  \pm  0.42$ &$    26.36 \pm   0.35 $&$0.433  $\\
J221941.85-074703.8 &$ 14.91 \pm 0.58$ &$     2.52  \pm   0.57$ &$     2.88  \pm  0.52$ &$     2.38 \pm   0.34 $&$0.369  $\\
J221939.28-074744.5 &$  1.72 \pm 0.37$ &$     0.88  \pm   0.14$ &$     0.31  \pm  0.34$ &$     0.62 \pm   0.32 $&$0.351  $\\
J221941.90-074825.7 &$ 29.10 \pm 0.60$ &$     7.12  \pm   0.49$ &$     6.26  \pm  0.63$ &$    15.38 \pm   0.21 $&$0.375  $\\

\noalign{\smallskip}
\hline
\noalign{\bigskip}
\label{tab:values}
\end{tabular}
\end{small} \\
\begin{minipage}{14.5cm}
\begin{small}
NOTES: Col. (1): SDSS name of the galaxy.  Col.(2-5): Equivalent
widths of the \od\/, \Hb\/, \oiiiq\/, and \oiiic\/ emission lines,
respectively.  Col.(6): Measured redshift.
\end{small}
\end{minipage}
\end{table*}
%%%%%%%%%%%%%%%%%%%%%%%%%%%%%%%%%%%%%%%%%%%%%%%%%%%%%%%%%%%%%

\section{Data analysis and gas-phase metallicity}

The apparent $g$-band Petrosian magnitudes of the sample galaxies
range between $24.9$ and $ 20.8 {\rm~mag}$, as they were derived from
the SDSS DR7. Since the $S/N$ of the spectra was too low to derive the
radial distribution of the gas metallicity, we rebinned the spectra
along the spatial direction to obtain a $S/N \ge 15$ per resolution
element.

\begin{figure}
\includegraphics[width=0.5\textwidth]{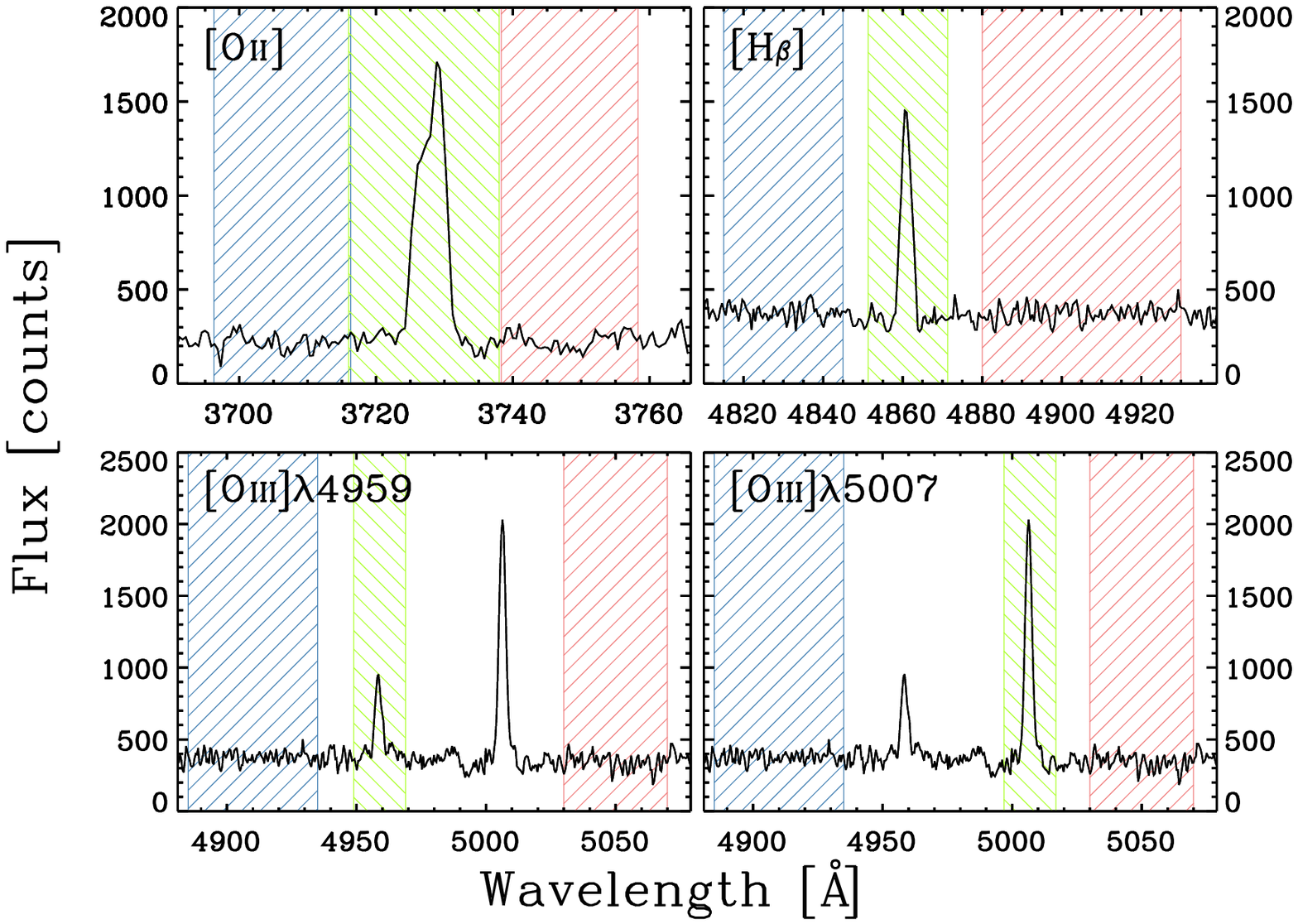}
\caption{Central (green dashed region) and continuum bandpasses (blue
  and red dashed regions) adopted for measuring the equivalent widths
  of the \od\/ (top left panel), \Hb\/ (top right panel), \oiiiq\/
  (bottom left panel), and \oiiic\/ (bottom right panel) emission
  lines. }
\label{ew}
\end{figure}

 Since no spectrophotometric standard star had been observed, the
  galaxy spectra were not flux-calibrated. Furthermore, the observed
  spectral range does not include the \Ha\/ emission line and thus a
  reliable reddening estimation was not possible. For the above
  reasons, we followed the approach suggested by \cite{kobulnicky2003}
  by replacing the flux of the emission-lines with their equivalent
  width when measuring the gas-phase metallicity. This method does not
  need flux-calibrated spectra and has the further advantage of being
  insensitive to reddening. In addition, \cite{kobulnicky2003}
  verified that the error associated with the use of the emission-line
  equivalent widths is smaller than both the typical error in the flux
  measurement and systematic error in the \rdt\/ and \oh\/
  calibrations.
Thus, we measured the equivalent widths of the {\od\/}, \Hb\/, and
{\ot\/} emission lines in the-rest frame spectra.
For each emission line, we considered a central bandpass covering the
feature of interest and two adjacent bandpasses, at the red and blue
side, tracing the local continuum. The continuum level underlying the
emission line was estimated by interpolating a straight line in the
continuum bandpasses. The bandpasses were defined following
\citet{fisher1998} for \od\/ and \citet{tesigonzalez} for \Hb\/,
\oiiiq\/, and \oiiic\/. The bandpasses are listed in Table
\ref{tab:ewregions} and shown in Figure \ref{ew}. The errors
associated with the measured equivalent widths were derived from
photon statistics and CCD read-out noise, and calibrated by means of
Monte Carlo simulations.

 A well-known problem in measuring the equivalent width of the
  \Hb\ emission line in galaxies is the contamination by the \Hb\/
  absorption line. We tried to address this issue by using GANDALF
  (Gas AND Absorption Line fitting; \citealt{sarzetal06}) to fit the
  galaxy spectra with synthetic stellar population models, as done by
  \citet{moreetal08, moreetal12}. A linear combination of template
  stellar spectra from the ELODIE library by \citet{prusou01} was
  convolved with the line-of-sight velocity distribution and fitted to
  the observed galaxy spectrum by $\chi^2$ minimization in pixel
  space. However, the low $S/N$ of the stellar continuum of the galaxy
  spectra prevented us from deriving a reliable combination of stellar
  templates to fit the galaxy stellar component. Therefore, to account
  for the \Hb\ absorption-line contamination we followed
  \cite{kobulnicky2004} by applying a correction of 2 $\AA$ \citep{kobulnicky2003} to the
  measured equivalent width of the \Hb\/ emission line used to
  derive the gas-phase metallicity.

All the equivalent widths measured for the galaxies in our sample are
reported in Table \ref{tab:values}.  They were used to compute the
metallicity-sensitive emission-line ratio \rdt, as well as the
emission-line ratio \otd\/ defined by \citet{kobulnicky2004} as
\begin{equation}
{{\it O}_{32}}=\frac{[\ion{O}{iii}]\lambda4959+[\ion{O}{iii}]\lambda5007},
{[\ion{O}{ii}]\lambda3727}
\end{equation}
which is mostly sensitive to the ionization \citep{nagao2006}.  All the
values of log(\rdt) and log(\otd) are listed in Table
\ref{tab:values_met}.

%%%%%%%%%%%%%%%%%%%%%%%%%%%%%%%%%%%%%%%%%%%%%%%%%%%%%%%%%%%%%
%% TABELLA BANDE
\begin{table}
\caption[EW intervals definition]{Definition of the bandpasses for the emission lines.}
\centering
\begin{tiny}
\begin{tabular}{lrrc}
\hline
\noalign{\smallskip}
\multicolumn{1}{c}{Emission line} &
\multicolumn{1}{c}{Central bandpass} &
\multicolumn{1}{c}{Continuum bandpasses} &
\multicolumn{1}{c}{Reference}  \\
\multicolumn{1}{c}{ } &
\multicolumn{1}{c}{[$\AA$]} &
\multicolumn{1}{c}{[$\AA$]} &
\multicolumn{1}{c}{} \\
\multicolumn{1}{c}{(1)} &
\multicolumn{1}{c}{(2)} &
\multicolumn{1}{c}{(3)} &
\multicolumn{1}{c}{(4)} \\
\noalign{\smallskip}
\hline
\noalign{\smallskip}
\od        & $3716.30 - 3738.30$ & $3696.30 - 3716.30$ & 1 \\
           &                     & $3738.30 - 3758.30$ & 1 \\
\Hb        & $4851.32 - 4871.32$ & $4815.00 - 4845.00$ & 2\\
           &                     & $4880.00 - 4930.00$ & 2\\
\oiiiq     & $4948.92 - 4968.92$ & $4885.00 - 4935.00$ & 2\\
           &                     & $5030.00 - 5070.00$ & 2\\
\oiiic     & $4996.85 - 5016.85$ & $4885.00 - 4935.00$ & 2\\
           &                     & $5030.00 - 5070.00$ & 2\\
\noalign{\smallskip}
\hline
\noalign{\bigskip}
\label{tab:ewregions}
\end{tabular}
\end{tiny}
\begin{minipage}{18.5cm}
\begin{small}
NOTES: Col.(4): 1 = \cite{fisher1998}, 2 = \citet{tesigonzalez}
\end{small}
\end{minipage}
\end{table}
%%%%%%%%%%%%%%%%%%%%%%%%%%%%%%%%%%%%%%%%%%%%%%%%%%%%%%%%%%%%%

%%%%%%%%%%%%%%%%%%%%%%%%%%%%%%%%%%%%%%%%%%%%%%%%%%%%%%%%%%%%%
%% FIGURA BANDE
%%%%%%%%%%%%%%%%%%%%%%%%%%%%%%%%%%%%%%%%%%%%%%%%%%%%%%%%%%%%%

For the sample galaxies, we could not derive the \nii\ $/$\Ha\/ ratio,
thus it was impossible to break the \rdt\/ degeneracy. Following
\citet{Kobulnicky1999} and \citet{Kobulnickyetal2003}, we assumed that all
the galaxies lie in the upper branch (\oh$\geq\,8.4$) of the
\rdt\/-O/H relation. We then checked the consistency of this
  assumption by adopting the [\ion{O}{iii}]/[\ion{O}{ii}] diagnostic lines
  ratio as proposed by \citet{maiolino2008}.  As done by
\citet{kobulnicky2004}, we derived the gas-phase oxygen abundance by
averaging the values obtained from the calibrations for the upper
branch given in \citet{mcgaugh1991} and \citet{kewleydopita2002}. The
values of \oh\/ and their uncertainties, owing to the statistical
measurement errors in the equivalent widths, are listed in Table
\ref{tab:values_met}. The systematic errors due to the uncertainties
in the photoionization models \citep[$0.2-0.5$ dex;][]{Kennicutt2003,
  garnet2004} were not taken into account.

\begin{figure}[htbp!]
\includegraphics[width=0.5\textwidth]{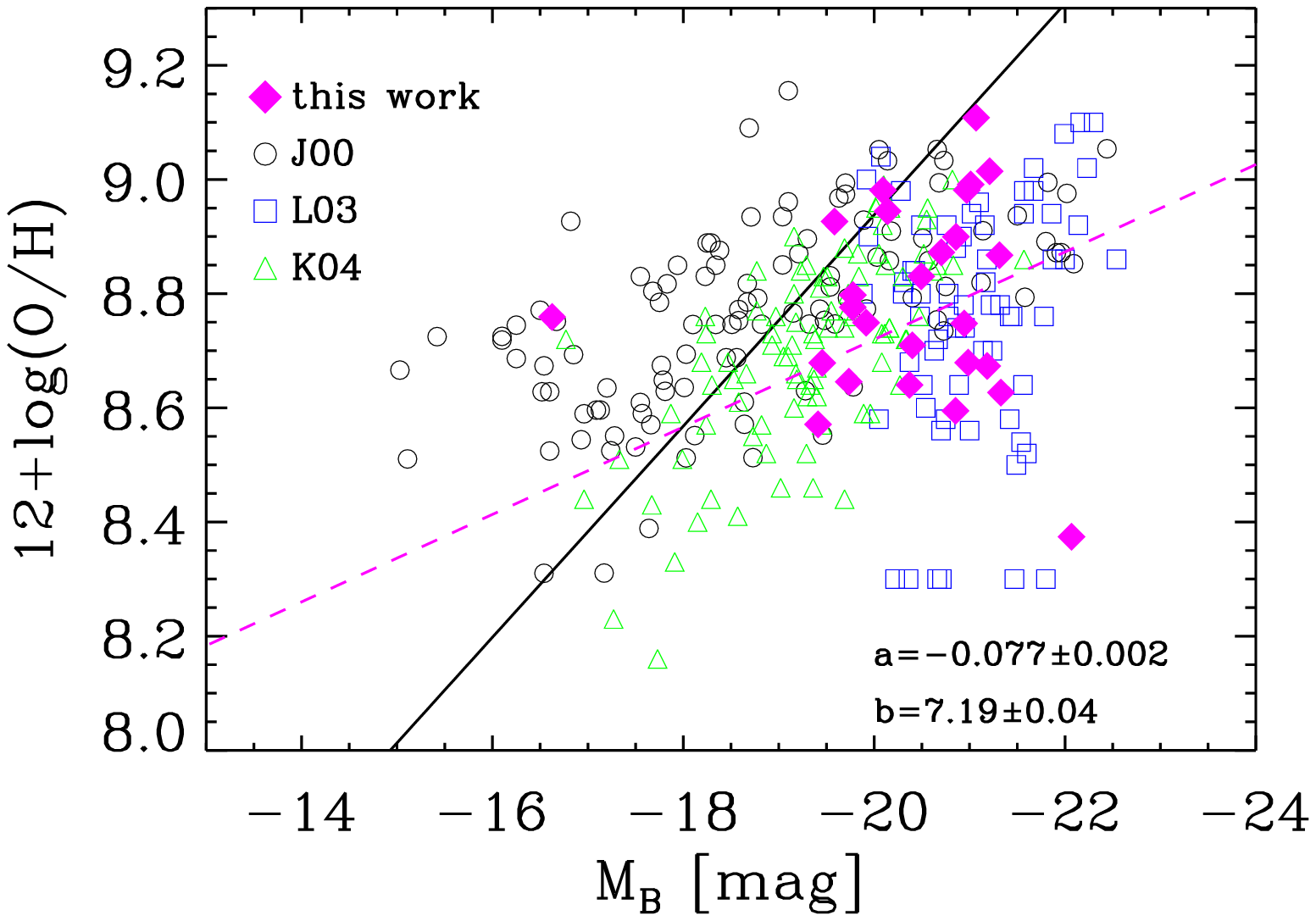}
\caption{Plot of our data (solid diamonds) on the $L-Z$ plane. Open
  squares represent the galaxies in the redshift range $0.47< z <
  0.92$ from the Canada-France Redshift Survey \citep{lilly2003}, open
  triangles mark the galaxies with $0.3< z < 0.6$ from the GOODS-North
  Field \citep{kobulnicky2004}, and open circles refer to the local
  galaxies of the Nearby Field Galaxy Sample \citep{jansen2000}. The
  solid line indicates the $L-Z$ relation obtained by
  \citet{tremonti2004} for local galaxies. The magenta dashed
    line represents the linear fit (\oh = a$\cdot M_{\rm B}$ + b)
    through our data points except for J154925.24-003717.8 and
    J003806.59+000421.9.} 
\label{fig:L-Z}
\end{figure}

In Figure \ref{fig:L-Z}, we plot the oxygen abundance as a function
of the absolute magnitude in the $B$-band $M_{\rm B}$ for the sample
galaxies.  We derived $M_{\rm B}$ from the rest-frame $g$-band
Petrosian magnitude given in the SDSS DR7, using the transformation
$B\,=\,g\,+\,0.327\,(g-r)\,+\,0.216$ \citep{chogas08}. The
  resulting values are given in Table \ref{tab:values_met}. We also
plotted the data for local (Nearby Field Galaxy Sample,
\citealt{jansen2000}; SDSS, \citealt{tremonti2004}) and
intermediate-redshift galaxies (Canada-France Redshift Survey,
\citealt{lilly2003}; GOODS-North Field, \citealt{kobulnicky2004}).

The dashed magenta line in Figure \ref{fig:L-Z} represents the linear
fit to the sample galaxies.  J154925.24-003717.8 and
J003806.59+000421.9 are two outliers in the $L-Z$ distribution. The
most prominent outlier is the faintest galaxy of the sample, for which
the assumption of an upper \rdt\/-O/H branch could be incorrect
\citep{skillman89}. Adopting the calibration of
\citet{kewleydopita2002} for the lower branch, we obtained a value of
\oh\/$=8.14$. However only the measurement of the \nii\ $/$\Ha\/ ratio
could firmly help us to discriminate between the two branches. The second outlier
has an extremely low value of the gas-phase metallicity. This is
probably due to an underestimation of the \Hb\ absorption-line
correction. The slope and zero-point of the fitted $L-Z$ relation to
our data points (calculated after excluding the two outliers) are
consistent within the errors with the slope and zero-point obtained by
\cite{kobulnicky2004} for their sample of galaxies in the redshift
range $0.2-0.4$. This was expected since 23 of our 27 galaxies
have a redshift in the range $0.35-0.45$, and confirms their findings of an
evolution in the $L-Z$ relation since intermediate redshift.

Finally, we estimated the star formation rate (SFR) of the sample
  galaxies from their $B-$band absolute magnitudes and \Hb\/
  equivalent widths according to the relations of
  \citet{Kennicutt2003} and \citet{kobulnicky2004}
\begin{equation}
{\rm SFR~(M_{\odot} yr^{-1})}=\frac{2.8 \times 5.49 \cdot 10^{31} \times 2.5^{M_{\rm B}} \cdot EW_{{\rm H}{\small{\beta}}}}{1.26 \cdot 10^{41}}.
\end{equation}
The derived SFRs are reported in Table \ref{tab:values_met}.

%%%%%%%%%%%%%%%%%%%%%%%%%%%%%%%%%%%%%%%%%%%%%%%%%%%%%%%%%%%%%
%% TABELLA VALORI DERIVATI
\begin{table*}
\centering
\caption{Derived properties of the sample galaxies.}
\begin{tiny}
\begin{tabular}{crrrrr}
\hline
\noalign{\smallskip}
\multicolumn{1}{c}{Galaxy name} &
\multicolumn{1}{c}{$M_{\rm B}$} &
\multicolumn{1}{c}{log(\rdt)} &
\multicolumn{1}{c}{log(\otd)} &
\multicolumn{1}{c}{\oh} &
\multicolumn{1}{c}{SFR} \\
\multicolumn{1}{c}{SDSS} &
\multicolumn{1}{c}{[mag]} &
\multicolumn{1}{c}{} &
\multicolumn{1}{c}{} &
\multicolumn{1}{c}{} &
\multicolumn{1}{c}{[M$_{\odot}$ yr$^{-1}$]} \\
\multicolumn{1}{c}{(1)} &
\multicolumn{1}{c}{(2)} &
\multicolumn{1}{c}{(3)} &
\multicolumn{1}{c}{(4)} &
\multicolumn{1}{c}{(5)} &
\multicolumn{1}{c}{(6)} \\
\noalign{\smallskip}
\hline
\noalign{\smallskip}
J003845.31+000006.8 &$ -22.06 $&  $ 1.06^{+0.01}_{-0.01} $&$   1.09^{+0.06}_{-0.07} $&$8.37 \pm 0.03$&$ 1.05 $\\
J003838.68+000225.1 &$ -21.32 $&  $ 0.80^{+0.02}_{-0.02} $&$  -0.17^{+0.01}_{-0.02} $&$8.63 \pm 0.02$&$ 8.33 $\\
J003806.59+000421.9 &$ -21.46 $&  $ 1.16^{+0.01}_{-0.01} $&$  -0.62^{+0.02}_{-0.03} $&$7.72 \pm 0.06$&$ 6.71 $\\
J134235.94+014424.1 &$ -20.09 $&  $ 0.40^{+0.02}_{-0.02} $&$   0.44^{+0.05}_{-0.06} $&$8.98 \pm 0.02$&$ 0.45 $\\
J134229.37+015244.5 &$ -19.91 $&  $ 0.72^{+0.01}_{-0.01} $&$   0.02^{+0.01}_{-0.03} $&$8.75 \pm 0.01$&$ 2.11 $\\
J134156.84+014241.0 &$ -19.58 $&  $ 0.53^{+0.02}_{-0.02} $&$   0.44^{+0.04}_{-0.04} $&$8.93 \pm 0.02$&$ 0.36 $\\
J145221.51+043448.8 &$ -20.97 $&  $ 0.40^{+0.01}_{-0.01} $&$  -0.65^{+0.04}_{-0.05} $&$8.98 \pm 0.03$&$ 2.46 $\\
J145150.75+044111.6 &$ -21.01 $&  $ 0.38^{+0.03}_{-0.03} $&$  -0.10^{+0.06}_{-0.08} $&$8.99 \pm 0.04$&$ 0.90 $\\
J145147.24+043654.2 &$ -20.94 $&  $ 0.69^{+0.02}_{-0.02} $&$  -0.45^{+0.06}_{-0.08} $&$8.75 \pm 0.06$&$ 1.65 $\\
J145156.20+043321.7 &$ -21.18 $&  $ 0.81^{+0.01}_{-0.01} $&$   0.38^{+0.03}_{-0.03} $&$8.67 \pm 0.02$&$ 0.70 $\\
J145147.77+043244.5 &$ -20.36 $&  $ 0.79^{+0.02}_{-0.03} $&$  -0.23^{+0.06}_{-0.08} $&$8.64 \pm 0.07$&$ 0.16 $\\
J145149.20+043501.6 &$ -19.77 $&  $ 0.68^{+0.02}_{-0.02} $&$  -0.15^{+0.05}_{-0.06} $&$8.78 \pm 0.05$&$ 0.26 $\\
J152638.18+034353.8 &$ -20.14 $&  $ 0.48^{+0.01}_{-0.01} $&$   0.11^{+0.02}_{-0.02} $&$8.94 \pm 0.01$&$ 1.23 $\\
J152630.18+035110.8 &$ -20.85 $&  $ 0.53^{+0.02}_{-0.02} $&$  -0.40^{+0.06}_{-0.07} $&$8.90 \pm 0.04$&$ 1.44 $\\
J152559.83+035249.3 &$ -20.98 $&  $ 0.76^{+0.01}_{-0.01} $&$  -0.19^{+0.03}_{-0.03} $&$8.68 \pm 0.03$&$ 0.88 $\\
J152606.03+034530.7 &$ -19.73 $&  $ 0.80^{+0.03}_{-0.02} $&$   0.02^{+0.04}_{-0.01} $&$8.64 \pm 0.01$&$ 0.78 $\\
J154933.50-004255.0 &$ -19.45 $&  $ 0.79^{+0.01}_{-0.02} $&$   0.22^{+0.01}_{-0.01} $&$8.68 \pm 0.02$&$ 1.00 $\\
J154931.48-003406.1 &$ -19.41 $&  $ 0.89^{+0.02}_{-0.03} $&$   0.47^{+0.02}_{-0.02} $&$8.57 \pm 0.03$&$ 1.38 $\\
J154925.24-003717.8 &$ -16.62 $&  $ 0.71^{+0.01}_{-0.01} $&$  -0.04^{+0.01}_{-0.01} $&$8.76 \pm 0.01$&$ 0.02 $\\
J154914.49-004010.2 &$ -20.49 $&  $ 0.62^{+0.01}_{-0.01} $&$  -0.29^{+0.02}_{-0.03} $&$8.83 \pm 0.02$&$ 1.57 $\\
J154925.12-003645.4 &$ -20.70 $&  $ 0.58^{+0.01}_{-0.01} $&$  -0.28^{+0.02}_{-0.02} $&$8.87 \pm 0.01$&$ 1.16 $\\
J222013.98-074355.8 &$ -21.21 $&  $ 0.33^{+0.02}_{-0.03} $&$  -0.44^{+0.07}_{-0.09} $&$9.01 \pm 0.05$&$ 0.69 $\\
J222014.86-073333.7 &$ -21.31 $&  $ 0.58^{+0.02}_{-0.03} $&$  -0.30^{+0.07}_{-0.09} $&$8.87 \pm 0.06$&$ 2.62 $\\
J221934.90-073825.9 &$ -20.85 $&  $ 0.89^{+0.01}_{-0.01} $&$   0.65^{+0.04}_{-0.05} $&$8.59 \pm 0.02$&$ 0.83 $\\
J221941.85-074703.8 &$ -19.77 $&  $ 0.65^{+0.01}_{-0.01} $&$  -0.45^{+0.05}_{-0.05} $&$8.80 \pm 0.04$&$ 0.22 $\\
J221939.28-074744.5 &$ -21.06 $&  $ 0.04^{+0.08}_{-0.09} $&$  -0.26^{+0.19}_{-0.34} $&$9.11 \pm 0.13$&$ 0.25 $\\
J221941.90-074825.7 &$ -20.39 $&  $ 0.74^{+0.01}_{-0.01} $&$  -0.12^{+0.02}_{-0.01} $&$8.71 \pm 0.01$&$ 1.13 $\\
\noalign{\smallskip}
\hline
\noalign{\bigskip}
\label{tab:values_met}
\end{tabular}
\end{tiny}
\begin{minipage}{14.5cm}
\begin{small}
NOTES:Col.(2): Absolute $B$-band magnitude from SDSS DR7
derived adopting the measured redshift. Col.(3): Metallicity-sensitive \rdt\/ parameter. Col.(4):
Ionization-sensitive \otd\/ parameter. Col.(5): Gas-phase oxygen
abundance. Col.(6): Star formation rate as derived using the \Hb\/ 
uncorrected for  \Hb\/ stellar absorption line.
\end{small}
\end{minipage}
\end{table*}
%%%%%%%%%%%%%%%%%%%%%%%%%%%%%%%%%%%%%%%%%%%%%%%%%%%%%%%%%%%%%

%%%%%%%%%%%%%%%%%%%%%%%%%%%%%%%%%%%%%%%%%%%%%%%%%%%%%%%%%%%%%
%%%%%%%%%%%%%%%%%%%%%%%%%%%%%%%%%%%%%%%%%%%%%%%%%%%%%%%%%%%%%%%%%%%%%%%%

\section{Conclusion}

Our original purpose was to explore the evolution of the $L-Z$ relation
of \cite{tremonti2004} to the intermediate-redshift range using a
large number of background galaxies in very deep VIMOS data.  After
combining all the constraints needed to perform our analysis, the
number of objects dramatically decreased from 1347 to 27, preventing us
from being able to draw any conclusion about the possible evolution of the $L-Z$ relation
with redshift. Here, we have reported the measured equivalent widths of the
\od\/, \Hb\/, \oiiiq, and \oiiic\/ emission lines, the values of the
metallicity-sensitive emission-line ratio \rdt, the ionization-sensitive
emission-line ratio \otd\/, and the gas-phase oxygen abundances \oh\/
derived for the serendipitous sample of 27 galaxies with the aim of
making them available to the community. Our measurements are
consistent with those of \citet{lilly2003} and \citet{kobulnicky2004}
for galaxies in the redshift range $0.3\,<\,z\,<\,1.0$.

\begin{acknowledgements}
We thank the anonymous referee for comments that improved the completeness
of this paper. This work was supported by Padua University through the
grants CPDA089220/08, 60A02-5934/09, and 60A02-1283/10 and by the Italian
Space Agency through the grant ASI-INAF I/009/10/0.  L.M. acknowledges
financial support from Padua University grant CPS0204. V. C. is
grateful to STScI for hospitality during the preparation of this
paper.

\end{acknowledgements}
%%%%%%%%%%%%%%%%%%%%%%%%%%%%%%%%%%%%%%%%%%%%%%%%%%%%%%%%%%%%%%%%%%
%%%%%%%%%%%%%%%%%%%%%%%%%%%%%%%%%%%%%%%%%%%%%%%%%%%%%%%%%%%%%%%%%%

\bibliographystyle{aa} % style aa.bst
\bibliography{metb} % your references Yourfile.bib

\begin{thebibliography}{45}
\expandafter\ifx\csname natexlab\endcsname\relax\def\natexlab#1{#1}\fi

\bibitem[{{Abazajian} {et~al.}(2009){Abazajian}, {Adelman-McCarthy},
  {Ag{\"u}eros}, {Allam}, {Allende Prieto}, {An}, {Anderson}, {Anderson},
  {Annis}, {Bahcall}, \& et~al.}]{DR7}
{Abazajian}, K.~N., {Adelman-McCarthy}, J.~K., {Ag{\"u}eros}, M.~A., {et~al.}
  2009, \apjs, 182, 543

\bibitem[{{Carollo} \& {Lilly}(2001)}]{carollo2001}
{Carollo}, C.~M. \& {Lilly}, S.~J. 2001, \apjl, 548, L153

\bibitem[{{Chonis} \& {Gaskell}(2008)}]{chogas08}
{Chonis}, T.~S. \& {Gaskell}, C.~M. 2008, \aj, 135, 264

\bibitem[{{Cresci} {et~al.}(2012){Cresci}, {Mannucci}, {Sommariva}, {Maiolino},
  {Marconi}, \& {Brusa}}]{cresci2012}
{Cresci}, G., {Mannucci}, F., {Sommariva}, V., {et~al.} 2012, \mnras, 421, 262

\bibitem[{{Dalcanton}(2007)}]{dalcanton2007}
{Dalcanton}, J.~J. 2007, \apj, 658, 941

\bibitem[{{Edmunds} \& {Pagel}(1984)}]{edmundspagel1984}
{Edmunds}, M.~G. \& {Pagel}, B.~E.~J. 1984, \mnras, 211, 507

\bibitem[{{Erb} {et~al.}(2006){Erb}, {Shapley}, {Pettini}, {Steidel}, {Reddy},
  \& {Adelberger}}]{erb2006}
{Erb}, D.~K., {Shapley}, A.~E., {Pettini}, M., {et~al.} 2006, \apj, 644, 813

\bibitem[{{Fisher} {et~al.}(1998){Fisher}, {Fabricant}, {Franx}, \& {van
  Dokkum}}]{fisher1998}
{Fisher}, D., {Fabricant}, D., {Franx}, M., \& {van Dokkum}, P. 1998, \apj,
  498, 195

\bibitem[{{Gallazzi} {et~al.}(2005){Gallazzi}, {Charlot}, {Brinchmann},
  {White}, \& {Tremonti}}]{gallazzi2005}
{Gallazzi}, A., {Charlot}, S., {Brinchmann}, J., {White}, S.~D.~M., \&
  {Tremonti}, C.~A. 2005, \mnras, 362, 41

\bibitem[{{Garnett} {et~al.}(2004){Garnett}, {Kennicutt}, \&
  {Bresolin}}]{garnet2004}
{Garnett}, D.~R., {Kennicutt}, Jr., R.~C., \& {Bresolin}, F. 2004, \apjl, 607,
  L21

\bibitem[{{Garnett} {et~al.}(1997){Garnett}, {Shields}, {Skillman}, {Sagan}, \&
  {Dufour}}]{garnett1997}
{Garnett}, D.~R., {Shields}, G.~A., {Skillman}, E.~D., {Sagan}, S.~P., \&
  {Dufour}, R.~J. 1997, \apj, 489, 63

\bibitem[{{Gonzalez}(1993)}]{tesigonzalez}
{Gonzalez}, J.~J. 1993, PhD thesis, Univ. California, Santa Cruz

\bibitem[{{Jansen} {et~al.}(2000){Jansen}, {Fabricant}, {Franx}, \&
  {Caldwell}}]{jansen2000}
{Jansen}, R.~A., {Fabricant}, D., {Franx}, M., \& {Caldwell}, N. 2000, \apjs,
  126, 331

\bibitem[{{Kennicutt} {et~al.}(2003){Kennicutt}, {Bresolin}, \&
  {Garnett}}]{Kennicutt2003}
{Kennicutt}, Jr., R.~C., {Bresolin}, F., \& {Garnett}, D.~R. 2003, \apj, 591,
  801

\bibitem[{{Kewley} \& {Dopita}(2002)}]{kewleydopita2002}
{Kewley}, L.~J. \& {Dopita}, M.~A. 2002, \apjs, 142, 35

\bibitem[{{Kobulnicky} \& {Kewley}(2004)}]{kobulnicky2004}
{Kobulnicky}, H.~A. \& {Kewley}, L.~J. 2004, \apj, 617, 240

\bibitem[{{Kobulnicky} \& {Phillips}(2003)}]{kobulnicky2003}
{Kobulnicky}, H.~A. \& {Phillips}, A.~C. 2003, \apj, 599, 1031

\bibitem[{{Kobulnicky} {et~al.}(2003){Kobulnicky}, {Willmer}, {Phillips},
  {Koo}, {Faber}, {Weiner}, {Sarajedini}, {Simard}, \&
  {Vogt}}]{Kobulnickyetal2003}
{Kobulnicky}, H.~A., {Willmer}, C.~N.~A., {Phillips}, A.~C., {et~al.} 2003,
  \apj, 599, 1006

\bibitem[{{Kobulnicky} \& {Zaritsky}(1999)}]{Kobulnicky1999}
{Kobulnicky}, H.~A. \& {Zaritsky}, D. 1999, \apj, 511, 118

\bibitem[{{Lara-L{\'o}pez} {et~al.}(2010){Lara-L{\'o}pez}, {Bongiovanni},
  {Cepa}, {P{\'e}rez Garc{\'{\i}}a}, {S{\'a}nchez-Portal}, {Casta{\~n}eda},
  {Fern{\'a}ndez Lorenzo}, \& {Povi{\'c}}}]{laralopez2010}
{Lara-L{\'o}pez}, M.~A., {Bongiovanni}, A., {Cepa}, J., {et~al.} 2010, \aap,
  519, A31

\bibitem[{{Lilly} {et~al.}(2003){Lilly}, {Carollo}, \& {Stockton}}]{lilly2003}
{Lilly}, S.~J., {Carollo}, C.~M., \& {Stockton}, A.~N. 2003, \apj, 597, 730

\bibitem[{{Maier} {et~al.}(2005){Maier}, {Lilly}, {Carollo}, {Stockton}, \&
  {Brodwin}}]{maier2005}
{Maier}, C., {Lilly}, S.~J., {Carollo}, C.~M., {Stockton}, A., \& {Brodwin}, M.
  2005, \apj, 634, 849

\bibitem[{{Maiolino} {et~al.}(2008){Maiolino}, {Nagao}, {Grazian}, {Cocchia},
  {Marconi}, {Mannucci}, {Cimatti}, {Pipino}, {Ballero}, {Calura}, {Chiappini},
  {Fontana}, {Granato}, {Matteucci}, {Pastorini}, {Pentericci}, {Risaliti},
  {Salvati}, \& {Silva}}]{maiolino2008}
{Maiolino}, R., {Nagao}, T., {Grazian}, A., {et~al.} 2008, \aap, 488, 463

\bibitem[{{Mannucci} {et~al.}(2010){Mannucci}, {Cresci}, {Maiolino}, {Marconi},
  \& {Gnerucci}}]{mannucci2010}
{Mannucci}, F., {Cresci}, G., {Maiolino}, R., {Marconi}, A., \& {Gnerucci}, A.
  2010, \mnras, 408, 2115

\bibitem[{{Mannucci} {et~al.}(2009){Mannucci}, {Cresci}, {Maiolino}, {Marconi},
  {Pastorini}, {Pozzetti}, {Gnerucci}, {Risaliti}, {Schneider}, {Lehnert}, \&
  {Salvati}}]{mannucci2009}
{Mannucci}, F., {Cresci}, G., {Maiolino}, R., {et~al.} 2009, \mnras, 398, 1915

\bibitem[{{McClure} \& {van den Bergh}(1968)}]{mcclure1968}
{McClure}, R.~D. \& {van den Bergh}, S. 1968, \aj, 73, 1008

\bibitem[{{McGaugh}(1991)}]{mcgaugh1991}
{McGaugh}, S.~S. 1991, \apj, 380, 140

\bibitem[{{Morelli} {et~al.}(2012){Morelli}, {Corsini}, {Pizzella}, {Dalla
  Bont{\`a}}, {Coccato}, {M{\'e}ndez-Abreu}, \& {Cesetti}}]{moreetal12}
{Morelli}, L., {Corsini}, E.~M., {Pizzella}, A., {et~al.} 2012, \mnras, 423,
  962

\bibitem[{{Morelli} {et~al.}(2008){Morelli}, {Pompei}, {Pizzella},
  {M{\'e}ndez-Abreu}, {Corsini}, {Coccato}, {Saglia}, {Sarzi}, \&
  {Bertola}}]{moreetal08}
{Morelli}, L., {Pompei}, E., {Pizzella}, A., {et~al.} 2008, \mnras, 389, 341

\bibitem[{{Moustakas} {et~al.}(2011){Moustakas}, {Zaritsky}, {Brown}, {Cool},
  {Dey}, {Eisenstein}, {Gonzalez}, {Jannuzi}, {Jones}, {Kochanek}, {Murray}, \&
  {Wild}}]{Moustakas2011}
{Moustakas}, J., {Zaritsky}, D., {Brown}, M., {et~al.} 2011, ArXiv e-prints
  1112.3300

\bibitem[{{Nagao} {et~al.}(2006){Nagao}, {Maiolino}, \& {Marconi}}]{nagao2006}
{Nagao}, T., {Maiolino}, R., \& {Marconi}, A. 2006, \aap, 459, 85

\bibitem[{{Osterbrock}(1989)}]{osterbrock1989}
{Osterbrock}. 1989, {Astrophysics of Gaseous Nebulae and Active Galactic
  Nuclei} (Mill Valley: University Science Books)

\bibitem[{{Pagel} {et~al.}(1979){Pagel}, {Edmunds}, {Blackwell}, {Chun}, \&
  {Smith}}]{pagel1979}
{Pagel}, B.~E.~J., {Edmunds}, M.~G., {Blackwell}, D.~E., {Chun}, M.~S., \&
  {Smith}, G. 1979, \mnras, 189, 95

\bibitem[{{Prugniel} \& {Soubiran}(2001)}]{prusou01}
{Prugniel}, P. \& {Soubiran}, C. 2001, \aap, 369, 1048

\bibitem[{{Richer} \& {McCall}(1995)}]{richer95}
{Richer}, M.~G. \& {McCall}, M.~L. 1995, \apj, 445, 642

\bibitem[{{Rodrigues} {et~al.}(2012){Rodrigues}, {Puech}, {Hammer}, {Rothberg},
  \& {Flores}}]{rodretal2012}
{Rodrigues}, M., {Puech}, M., {Hammer}, F., {Rothberg}, B., \& {Flores}, H.
  2012, \mnras, 421, 2888

\bibitem[{{Ryder}(1995)}]{ryder1995}
{Ryder}, S.~D. 1995, \apj, 444, 610

\bibitem[{{Sarzi} {et~al.}(2006){Sarzi}, {Falc{\'o}n-Barroso}, {Davies},
  {Bacon}, {Bureau}, {Cappellari}, {de Zeeuw}, {Emsellem}, {Fathi},
  {Krajnovi{\'c}}, {Kuntschner}, {McDermid}, \& {Peletier}}]{sarzetal06}
{Sarzi}, M., {Falc{\'o}n-Barroso}, J., {Davies}, R.~L., {et~al.} 2006, \mnras,
  366, 1151

\bibitem[{{Savaglio} {et~al.}(2005){Savaglio}, {Glazebrook}, {Le Borgne},
  {Juneau}, {Abraham}, {Chen}, {Crampton}, {McCarthy}, {Carlberg}, {Marzke},
  {Roth}, {Jorgensen}, \& {Murowinski}}]{savaglio2005}
{Savaglio}, S., {Glazebrook}, K., {Le Borgne}, D., {et~al.} 2005, \apj, 635,
  260

\bibitem[{{Shapley} {et~al.}(2004){Shapley}, {Erb}, {Pettini}, {Steidel}, \&
  {Adelberger}}]{shapley2004}
{Shapley}, A.~E., {Erb}, D.~K., {Pettini}, M., {Steidel}, C.~C., \&
  {Adelberger}, K.~L. 2004, \apj, 612, 108

\bibitem[{{Skillman} {et~al.}(1989){Skillman}, {Kennicutt}, \&
  {Hodge}}]{skillman89}
{Skillman}, E.~D., {Kennicutt}, R.~C., \& {Hodge}, P.~W. 1989, \apj, 347, 875

\bibitem[{{Sommariva} {et~al.}(2012){Sommariva}, {Mannucci}, {Cresci},
  {Maiolino}, {Marconi}, {Nagao}, {Baroni}, \& {Grazian}}]{sommariva2011}
{Sommariva}, V., {Mannucci}, F., {Cresci}, G., {et~al.} 2012, \aap, 539, A136

\bibitem[{{Tremonti} {et~al.}(2004){Tremonti}, {Heckman}, {Kauffmann},
  {Brinchmann}, {Charlot}, {White}, {Seibert}, {Peng}, {Schlegel}, {Uomoto},
  {Fukugita}, \& {Brinkmann}}]{tremonti2004}
{Tremonti}, C.~A., {Heckman}, T.~M., {Kauffmann}, G., {et~al.} 2004, \apj, 613,
  898

\bibitem[{{Yegorova} {et~al.}(2011){Yegorova}, {Pizzella}, \&
  {Salucci}}]{yegorova2011}
{Yegorova}, I.~A., {Pizzella}, A., \& {Salucci}, P. 2011, \aap, 532, A105

\bibitem[{{Zahid} {et~al.}(2011){Zahid}, {Kewley}, \& {Bresolin}}]{zahid2011}
{Zahid}, H.~J., {Kewley}, L.~J., \& {Bresolin}, F. 2011, \apj, 730, 137

\end{thebibliography}

%\begin{thebibliography}{}

\end{document}